\documentstyle[12pt]{article}
\begin{document}\baselineskip=18pt
\def\be{\begin{equation}}
\def\ee{\end{equation}}
\def\bearst{\begin{eqnarray*}}
\def\eearst{\end{eqnarray*}}
\def\peleven{\parbox{11cm}}
\def\peffec{\peight{\bearst\eearst}\hfill\peleven}
\def\pspace{\peight{\bearst\eearst}\hfill}
\def\ptwelve{\parbox{12cm}}
\def\peight{\parbox{8mm}}
\def\bear{\begin{eqnarray}}
\def\eear{\end{eqnarray}}
\def\E{{\rm e}}
\input epsf.tex
\newcommand{\slp}{\raise.15ex\hbox{$/$}\kern-.57em\hbox{$\partial$}}
\newcommand{\sla}{\raise.15ex\hbox{$/$}\kern-.57em\hbox{$a$}}
\newcommand{\slA}{\raise.15ex\hbox{$/$}\kern-.57em\hbox{$A$}}
\newcommand{\slB}{\raise.15ex\hbox{$/$}\kern-.57em\hbox{$B$}}
\newcommand{\slD}{\raise.15ex\hbox{$/$}\kern-.57em\hbox{$D$}}
\newcommand{\slb}{\raise.15ex\hbox{$/$}\kern-.57em\hbox{$b$}}
\newcommand{\slW}{\raise.15ex\hbox{$/$}\kern-.57em\hbox{$W$}}
\font\grrm=cmbx10 scaled 1200
\font\vb=cmbx10 scaled 1440
\font\bigcal=cmsy10 scaled 1200
\def\eightpoint{\def\rm{\fam0\eightrm}}
\def\flex{\raise 6pt\hbox{$\leftrightarrow $}\! \! \! \! \! \! }
\def\tr{ \mathop{\rm tr}}
\def\atanh{\mathop{\rm atanh}}
\def\Tr{\mathop{\rm Tr}}
\def\dal{\Box} 
\def\Natural{\hbox{\hskip 1.5pt\hbox to 0pt{\hskip -2pt I\hss}N}}
\def\Integer{\>\hbox{{\sf Z}} \hskip -0.82em \hbox{{\sf Z}}\,}
\def\Rational{\hbox{\hbox to 0pt{\hskip 2.7pt \vrule height 6.5pt
                                  depth -0.2pt width 0.8pt \hss}Q}}
\def\Real{\hbox{\hskip 1.5pt\hbox to 0pt{\hskip -2pt I\hss}R}}
\def\Complex{\hbox{\hbox to 0pt{\hskip 2.7pt \vrule height 6.5pt
                                  depth -0.2pt width 0.8pt \hss}C}}
\def \ln {{\rm ln}\, }
\def \cotg {\rm cotg }
\vskip 1cm
\begin{tabbing}
\hskip 11.5 cm \= \\
\>quant-ph/9803021\\
\>February 1998
\end{tabbing}
\vskip 1cm
\begin{center}
{\Large\bf Quantization of the multidimensional rotor}
\vskip 1.2cm
{\large \bf E. Abdalla$^a$ and 
R. Banerjee$^b$
\vskip 0.4cm
{{\it $^a$Instituto de F\'\i sica-USP,\\ C.P.66318, 05315-970
S. Paulo, Brazil}\\
\vskip 0.4cm
{\it $^b$S.N. Bose National Centre for Basic Sciences\\
Block JD, Sector III, Salt Lake City, Calcutta 700.091 India}}\\
{\sl $^a$ eabdalla@fma.if.usp.br\\
$^b$ rabin@boson.bose.res.in}}
\end{center}
\abstract 
We reconsider the problem of quantising a particle on the $D$-dimensional
sphere. Adopting a Lagrangian method of reducing the degrees of freedom,
the quantum Hamiltonian is found to be the usual Schr\"odinger 
operator without any boundary term. The equivalence with the Dirac
Hamiltonian approach is demonstrated, either in the cartesian or in the 
curvilinear basis. We also briefly comment on the path integral approach.

\vfill\eject
\section{Introduction and review of the problem}
The problem of the quantization of the rotor has been studied since
decades, but remains a subject of intense debate.

De Witt\cite{dewitt1,dewitt2} 
studied the path integral quantization, and proposed that a term proportional
to the curvature should be included in the Hamiltonian. However, problems 
connected to the definition of the path integral using curvilinear 
coordinates, unknown at that time, turns out to invalidate that proposal.
This is essentially tied to the fact that for quantisation in curvilinear 
coordinates, new terms may arise. Such happens even in the quantisation of free
particles in curvilinear coordinates. Edwards and Gulyaev\cite{edwards} 
carefully considered that problem, and showed that the quantization
via path integration may lead to different results, due to  problems
intrinsically connected to the path integral formulation\cite{rivers}. 
Though essential, this last paper has been frequently forgotten in the 
literature.

Dirac formulation\cite{dirac} 
of the problem has also been undertaken\cite{falck}. 
In this case one faces the question of ordering of the quantum operators, a 
question which can only be tackled using definite ordering prescriptions
based on general arguments as hermiticity and general coordinate invariance. 
The Laplace-Beltrami operator has been obtained using the Dirac quantization 
procedure without curvature terms in {\it e.g.} \cite{falck}, 
for the special case of a three-dimensional rotor. 

In the path integral formalism, besides the early developments\cite{edwards},
there are special definitions of the path integral on the surface
of the sphere which do not give rise to the curvature term\cite{kleibook}. 
A discrepancy with the Dirac formalism has however been reported in
several papers\cite{kleinert,girotti,marinov}, 
as well as other different results\cite{omotesato,asaa}. 

In spite of all these developments, the status of the problem is very 
confusing, and there have been many papers claiming a rejection of the Dirac
formalism\cite{kleinert} 
an intrinsic difference between path integral formulation
and operator formalism\cite{girotti}, 
or advocating different quantization schemes\cite{omotesato}. 
The importance of the problem can be appreciated from the fact
that it has implications in curved space quantization, when defining the
Wheeler-de-Witt equation\cite{orfeu}, 
and in the quantization of sigma model Lagrangians, important to string 
theories, where the Wheeler-de-Witt equation has central importance.

The aim of the present paper is to analyse the problem directly in terms of 
reduced coordinates by solving the constraint, thereby bypassing the 
ambiguities inherent in either the Dirac or path integral approaches. 
The classical reduced space is therefore obtained in a straightforward manner.
We then pass to the quantum formulation
by using the Laplace-Beltrami construction. It leads to a quantum Hamiltonian
which is the usual Schr\"odinger operator without any curvature term. The 
connection of our results with the Dirac approach is then established.
We conclude by briefly mentioning the path integral formulation,
where the counterpart of the above approach is the special time slicing of 
the integral as explained by T. D. Lee\cite{lee}. 

\section{Reduced coordinates}
The Lagrangian for a particle of unit mass constrained to move
on the surface of a $D$-dimensional sphere of radius $R$ is given by 
the well known expression
\be
{\cal L}= \frac 12 \dot x_\alpha\dot x_\alpha- \lambda 
\left(  x_\alpha x_\alpha -R^2
\right) \quad \alpha =1\cdots D\quad ,\label{lagrangian} 
\ee
where the constraint 
\be
\Omega =  x_\alpha x_\alpha -R^2 \approx 0\label{constraint} 
\ee
is implemented by the Lagrange multiplier $\lambda$.
Since this is a constrained system, the usual canonical approach must be
modified. A possible way is to use Dirac's constraint analysis. It
is known that this system presents second class constraints
so that the usual Poisson brackets have to be modified to the Dirac brackets.
Apart from the fact that these brackets are plagued by ordering ambiguities,
the extraction of the physical variables of the system is not very 
transparent. This has been analysed by several authors 
\cite{falck,kleinert,girotti,asaa}.

Here we shall adopt an alternative canonical approach developed recently
by one of us\cite{rb} which is based on a Lagrangian reduction by
systemmatically eliminating the unphysical variables using the equations of
constraint. From the constraint (\ref{constraint}) the coordinate
$x_D$ is expressed in terms of the remaining coordinates by
\be
x_D =\sqrt{R^2 -\vec x^2} \equiv \sqrt{R^2 - x_i^2},\quad i=1\cdots D-1
\label{solconstraint} 
\ee
Inserting this in equation (\ref{lagrangian}), we obtain the reduced
Lagrangian
\be
{\cal L}_r= \frac 12 g_{ij}\dot x_i\dot x_j\label{redlagrangian} 
\ee
where the metric is
\be
g_{ij}=\delta_{ij}+\frac { x_ix_j}{R^2-\vec x^2}\label{metric} 
\ee
Note that now the Lagrangian is expressed only in terms of the
uncostrained variables. Consequently, the conventional canonical
formalism is applicable. The canonical momenta are given by
\be
p_i =\frac {\partial {\cal L}_r}{\partial \dot x_i} = g_{ij}\dot x_j\quad .
\label{canomomenta} 
\ee
Since the system is uncostrained the velocities can be obtained
unambiguously by inverting (\ref{canomomenta}) to yield
\be
\dot x_i = g^{ij}p_j\label{velocities} 
\ee
where $ g^{ij}$ is the inverse of (\ref{metric}), being given by
\be
g^{ij}=\delta_{ij}-\frac { x_ix_j}{R^2}\label{inversemetric} 
\ee
The canonical Hamiltonian is now obtained by a Legendre transform
\be
{\cal H} = p_i\dot x_i -{\cal L}_r = \frac 12 p_i g^{ij} p_j
\label{hamiltonian} 
\ee
This gives the final expression for the classical reduced Hamiltonian.

In order to perform the quantization the above Hamiltonian is replaced by the
corresponding Laplace-Beltrami operator, being defined as
\be
\hat{\cal H}={\cal O}_{LB}=\frac 12 
g^{-1/4} \hat\pi_i g^{1/2} g^{ij}\hat\pi_j g^{-1/4}
\label{laplace-beltrami-hamil} 
\ee
where $\hat\pi$ is the quantum momentum operator,
and $g$ is the determinant of the metric $g_{ij}$. 

As is well known\cite{omotesato} the above transition from classical to quantum 
is based of hermiticity and general coordinate invariance. The quantum momentum 
operator satisfying these conditions is given by the expression
\be
\hat\pi_i=-i\hbar g^{-1/4}\vec\partial g^{1/4}\quad ,
\label{herm-momentum} 
\ee

An essential ingredient is the computation of the determinant of the
metric, which is sketched below,
\bear
g\equiv\det g_{ij}&=&\exp\tr\ln\left(\delta_{ij}+\frac { x_ix_j}{R^2-\vec x^2}
\right)\nonumber\\
&=&\exp\tr \frac{x_ix_j}{\vec x^2}\ln  \left( 1+\frac {\vec x^2}
{R^2-\vec x^2}\right) \label{determinant}\\ 
&=&\frac {R^2}{R^2-\vec x^2}\nonumber
\eear
It is now simple to obtain the quantum Hamiltonian,
\be
\hat {\cal H} =-\frac 12 \sqrt{R^2-\vec x^2}\vec\partial_i
\left( \delta_{ij}-\frac { x_ix_j}{R^2}\right) \left( {R^2-\vec x^2}
\right)^{-1/2} \vec\partial_j\, ,\quad \vec\partial_i=
\vec{\frac\partial{\partial x_i}}\quad . \label{operatorhamiltonian} 
\ee
The arrow above the derivative means that it operates on every
element on the right (i.e., it acts as a quantum operator).
It is now straightforward to show that the above operator is related
to the angular momentum in the reduced space,
\bear
L_{ij}&=&x_ip_j-x_jp_i =-i\hbar\left(x_i\partial_j-x_j\partial_i\right)
\quad ,\quad  \label{angularmomentumij}\\ 
L_{iD}&=&-\sqrt{R^2-\vec x^2}p_i=-L_{Di}=i\hbar
\sqrt{R^2-\vec x^2} \partial_i\quad ,
\label{angularmomentumid} 
\eear
Observe that since the $x_D$ coordinate has been eliminated, the conjugate
momentum $p_D$ does not exist in the reduced variables. Now it can
be verified that,
\be
\hat {\cal H}= \sum_{\alpha\beta}\frac  {L_{\alpha\beta}^2}{2 R^2}
\label{hamilangularmomenta} 
\ee
We thus find that the quantum Hamiltonian is the conventional
Schr\"odinger operator without any curvature term. This is the central result 
of the paper.

The above analysis was carried out in the cartesian basis, but it is 
instructive to repeat it in the curvilinear basis. Apart from serving as 
a consistency check, this will also illuminate the connection of the 
present study with the conventional Dirac approach. The mapping from the 
cartesian to the curvilinear coordinates is given by
\bear
x_D&=& r\cos\varphi_1 \nonumber\\
x_{D-1} &=& r\sin \varphi_1\cos \varphi_2\nonumber\\
x_{D-3} &=& r\sin\varphi_1\sin\varphi_2\cos\varphi_3\nonumber\\
\cdots &=& \cdots\label{x-curv}\\ 
x_2 &=& r\sin \varphi_1 \cdots \sin\varphi_{D-2}\cos \varphi_{D-1}\nonumber\\
x_1 &=& r\sin \varphi_1 \cdots \sin\varphi_{D-2}\sin \varphi_{D-1}\nonumber
\eear
In these variables, the Lagrangian is given by
\be
{\cal L}=\frac 12\lbrace \dot r^2 + r^2\dot\varphi_1^2 +r^2
\dot\varphi_2^2\sin^2\varphi_1 +\cdots  r^2 \dot\varphi_{D-1}^2 
\sin^2\varphi_1^2\cdots \sin^2\varphi_{D-2}\rbrace +\lambda 
\left( r-R\right) \label{kin-lag-curv} 
\ee
where the lagrange multiplier $\lambda$ defines the rotor constraints.
The canonical momenta are given by
\bear
\pi_\lambda &=&0\quad ,\quad \pi_r =\dot r\quad ,\nonumber\\
\pi_{\varphi_1} &=& r^2\dot\varphi_1\nonumber\\
\pi_{\varphi_2} &=& r^2\sin^2\varphi_1\;\dot
\varphi_2\nonumber\\
\cdots &=& \cdots \label{momentaconslagcurv}\\ 
\pi_{\varphi_D-1} &=& r^2\sin^2\varphi_1\cdots\sin^2\varphi_{n-1}\;
\dot\varphi_{D-1}\nonumber
\eear

In constrast to the cartesian analysis, the solution of the constraint
here is trivial and the reduced Hamiltonian obtsined by following the 
previous steps is
\be
{\cal H} =\frac 12 \sum g^{ab}\pi_a\pi_b\label{hamilgpipi} 
\ee
with 
\be
g^{ab} =\frac 1 {R^{2}} 
\left(\matrix{1 &0 &0 &\cdots &0\cr 
0 &\frac 1{\sin^2\varphi_1} & 0 &\cdots &0\cr
0 &0 &\frac 1 {\sin^2\varphi_1\sin^2\varphi_2} & \cdots & 0\cr
\cdots & \cdots & \cdots & \cdots & 0\cr
0 & 0 & 0      & \cdots & \frac 1 {\sin^2\varphi_1\sin^2\varphi_2\cdots 
\sin^2\varphi_{n-2}}\cr
}\right) 
\ee
The quantum Hamiltonian, as usual, is given by the corresponding 
Laplace-Beltrami operator. The momentum operator (\ref{herm-momentum}) 
is given by the expression
\be
\pi_{\varphi_i}=-i\hbar\frac 1
{\sin^{\frac{D-i}2}\varphi_i}\vec\partial_i\; 
{\sin^{\frac{D-i}2}\varphi_i}\quad , 
\ee
and in terms of the curvilinear variables for a constant radius the hamiltonian
operator (\ref{laplace-beltrami-hamil}) reduces to
\bear
\hat{\cal H} &=&-\frac{\hbar^2}{2R^2}
\sum_{i=1}^{n-1}\prod_{j=1}^{i-1}\frac 1{\sin^2\varphi_j}
\frac 1 {\left(\sin^{n-i-1}\varphi_i\right)^{n-i-1}}
\frac\partial{\partial
\varphi_i}\left(\sin\varphi_i\right)^{n-i-1}
\frac\partial{\partial \varphi_i}\nonumber\\
&\equiv &\frac{\hat L^2}{2R^2}\label{hamilislsquared-n} 
\eear
which in three dimensions simplifies to the well known expression
\be
\hat{\cal H} =-\frac{\hbar^2}{2R^2}\left(\frac 1 {\sin\theta}
\frac\partial{\partial
\theta}\sin\theta\frac\partial{\partial \theta} +\frac 1 {\left(\sin\theta
\right)^2}\frac{\partial^2}{\partial \varphi^2}\right)\equiv
\frac{\hat L^2}{2R^2}\label{hamilislsquared} 
\ee
which is the expected result and is compatible with our general expression
(\ref{hamilangularmomenta}) 
given in the cartesian basis.

To compare with the Dirac formalism we start from the original constrained 
Lagrangian in curvilinear coordinates given in (\ref{kin-lag-curv}). 
There is one primary constraint,
\be
\Omega = \pi_\lambda =0 \label{pi-lambda} 
\ee
To check for secondary constraints we construct the total Hamiltonian
\be
{\cal H}_T ={\cal H}_c + u\pi_\lambda 
\ee
where ${\cal H}_c$ is the canonical Hamiltonian given by
\be
{\cal H}_c = \frac 12 \lbrack \pi_r^2 +\frac{\pi_{\varphi_1}^2}{r^2}
+\frac{\pi_{\varphi_2}^2}{\sin^2\varphi_1 r^2} +\cdots +         
\frac{\pi_{\varphi_2}^2}{\prod_{j=1}^{n-2} \sin^2\varphi_j r^2} \rbrack
\label{cano-hamil-d-dim} 
\ee
Time conservation of the primary constraint leads to a secondary constraint.
Continuing this iterative process, the full set of constraints is
obtained.
\be
\Omega_1 =r-R\approx 0\quad ,\quad \Omega_2 =\pi_r \approx 0 
\ee
The unphysical canonical set $\lambda\, , \,\pi_\lambda$ associated with
the Lagrange multiplier is ignored. This leaves us with a pair of second
class constraints, $\Omega_1$ and $\Omega_2$. It is important
to point out that they form a canonical set,
\be
\lbrace \Omega_i,\Omega_j\rbrace =\epsilon _{ij} 
\ee
The special form of the constraints allows a straightforward
application of the Masakawa-Nakajima theorem\cite{mn} to extract the physical
variables and the Hamiltonian without the need of any explicit computation of
Dirac brackets. Using this theorem, it is simple to show that the canonical 
pairs are given by  $\varphi_i,\pi_{\varphi_i}$. 
In other words, the Dirac brackets among these variables
is equal to their Poisson brackets. The physical Hamiltonian, in terms 
of these pairs, is now obtained from the canonical Hamiltonian
by passing on to the constraint shell. This is found to coincide with the
reduced Hamiltonian (\ref{hamilgpipi}) obtained by our approach. The quantum 
Hamiltonian is then reobtained from the corresponding Laplace-Beltrami
operator.

The Dirac analysis of this problem in the cartesian basis is quite
nontrivial. This is essentially tied to the fact that the constraint algebra
is no longer canonical. The Dirac Brackets suffer from ordering problems
and the extraction of the canonical pairs of variables is quite 
non-trivial. It is precisely because of these reasons that Dirac analysis
has led to much confusion and 
controversy\cite{falck,kleinert,girotti}. 
However, by the Masakawa-Nakajima theorem, it is always possible to find the
canonical transformation which enables the extraction of the canonical
pair of variables without any further ambiguities. To keep our discussion
simple we consider below the example of the three dimensional rotator.

The constraint (\ref{constraint}) is given by
\be
\Omega_1 = x_ix_i -R^2=0\label{constraint-3d} 
\ee
The Hamiltonian following from a Legendre transform of the original
Lagrangian is simply given by
\be
{\cal H}= \frac 12 p_i^2 + \lambda\left( x_ix_i-R^2\right)
\label{hamil-3d-leg} 
\ee
which leads to the secondary constraint
\be
\Omega_2 = x_ip_i =0 \label{sec-constraint} 
\ee
This corresponds to the standard pair of constraints
$\left(\Omega_1,\Omega_2\right)$\cite{falck}. 
It is straightforward to compute the Dirac Brackets
\bear
\lbrace x_i,x_j\rbrace^* &=&0\nonumber\\
\lbrace x_i,p_j\rbrace^* &=&\delta_{ij} -\frac {x_ix_j}{R^2}
\label{dirac-bra}\\ 
\lbrace p_i,p_j\rbrace^* &=& -\frac 1 {R^2} 
\left(x_ip_j-x_jp_i\right)\nonumber
\eear
Note that the Dirac Brackets are plagued by ordering ambiguities. This
is bypassed by making a canonical transformation to a more convenient
set of variables. 
It is now simple to check that the transformation from cartesian to
the curvilinear coordinates given in (\ref{x-curv}) 
also defines the canonical transformation from the old to the new
set of variables,
\bear
&&x_1= R\sin\theta\cos\varphi\quad ,\quad x_2 =R\sin\theta\sin\varphi
\quad ,\quad x_3=R\cos\theta\nonumber\\
&& \pi_1 = \sin\theta\cos\varphi \pi_r +r\cos \theta\cos\varphi\pi_\theta
-r\sin\theta\sin\varphi\pi_\varphi\nonumber\\ 
&&\pi_2=\sin\theta\sin\varphi\pi_r+ r\cos \theta\sin\varphi\pi_\theta
+r\sin\theta\cos\varphi\pi_\varphi\nonumber\\
&&\pi_3=\cos \theta\pi_r-r\sin\theta\pi_\theta \label{pi-in-pirtphi} 
\eear
The canonical pairs are now well defined $(r,\pi_r)$ $(\theta,\pi_\theta)$
and $(\varphi,\pi_\varphi)$. The physical Hamiltonian is obtained
from ${\cal H}=\frac 12 p_ip_i$ using the canonical change of variables
and reads
\be
{\cal H}_{phys} = \frac 1{2R^2} \left( \pi_\theta^2 +\frac 1 {\sin^2\theta}
\pi_\varphi^2\right)\quad ,
\ee
where we have already pass on to the constraint shell,
\be
x_ip_i = r\pi_r =0\quad .
\ee
This completes the classical reduction. Since the above Hamiltonian 
is expressed in terms of canonical pairs, the Laplace-Beltrami 
construction goes through and we exactly reproduce the canonical
Hamiltonian (\ref{hamilislsquared}).

This completes our demonstration of the equivalence betwen the Lagrangian
reduction and the Hamiltonian reduction.

\section{Conclusions}

The main conclusion of the present work is that the quantum Hamiltonian
for the multidimensional rotor is given by the pure Schr\"odinger operator 
without any boundary term, provided we enforce the conditions of hermiticity 
and general coordinate invariance. This result was obtained in the 
Lagrangian formalism by directly solving the constraint and reducing the 
unwanted degrees of freedom. The Lagrangian analysis was done both in 
cartesian and in the curvilinear basis, following a technique recently 
suggested by one of us\cite{rb}. The equivalence 
with the Hamiltonian formalism of Dirac was also shown in either basis.

Some comments about the path integral are in order. The basic problem 
here stems from the fact that the definition of the path integral in 
curvilinear coordinates is 
rather tricky and subtle. It also appears in the original computation done by 
de Witt\cite{dewitt1} where a curvature term was found. However, the 
complications of working with path integrals with curvilinear basis was 
not appreciated at that time. This came to be highlighted only after the 
work of Edwards and  Gulyaev\cite{edwards}.

Indeed, an apparent clash between the canonical 
and (a naive) path integral formulation is already seen in the simplest of
examples, namely a free non relativistic particle in two 
dimensions\cite{kapoor}. Using the De Witt path integral prescription
it was shown that the free particle propagator obeys the following equation,
\be
i\hbar\frac\partial{\partial t}\psi =\left( -\frac{\hbar^2}2 \nabla^2
+\frac {\hbar^2}{8r^2}\right)\psi\label{schrodinger}
\ee
Surprisingly the above equation differs from the expected free
particle Schr\"odinger equation by an effective potential term.
The reason for this discrepancy is subtle. The computation in
\cite{kapoor} was done in polar coordinates, but using the naive
prescription of De Witt. However, as is well known by now,
the passage from the cartesian to the curvilinear basis introduces
curvature like terms. This has been explained lucidly in the textbook of 
Lee\cite{lee}. Indeed, for the particular problem at hand the explicit 
correction term has also been computed. It has been shown that the 
canonical Hamiltonian ${\cal H}_c$ cannot be used to define the path integral,
rather it must be ${\cal H}_c-\frac {\hbar^2}{8r^2}$\cite{lee}.
It is now obvious that with this modified Hamiltonian the correct
Schr\"odinger equation will be reproduced from (\ref{schrodinger}).
The lesson to be learnt is that the conventional De Witt path integral
prescription must be carefully applied for curvilinear coordinates.

Keeping these observations in mind, a clear computation as the one performed
by Kleinert\cite{kleibook} leads to the correct result. We however point 
out that a distinction can be made, so that it is possible to treat the 
particle either on
or near the surface of the sphere. For the former case the boundary term 
disappears\cite{kleibook} but in the latter, such a term exists. Since in the 
present analysis the constraints are always strongly enforced, we are 
confined to the case of the particle exactly on the sphere. A recent 
calculation from a more mathematical point of view also confirms our
result\cite{gh}. 

We conclude therefore that there is no clash between the canonical (either
Lagrangian or Hamiltonian) approach and the path integral formalism. Moreover,
the Dirac analysis also gives perfectly valid results, as elucidated here.

{\bf Acknowledgements}: this work has been partially supported
by Conselho Nacional de Desenvolvimento Cient\'\i fico e Tecnol\'ogico,
CNPq, Brazil and Funda\c c\~ao de Amparo a Pesquisa do Estado de
S\~ao Paulo (FAPESP), S\~ao Paulo, Brazil. One of us (E.A.) wishes to
thank the S.N. Bose National Centre for Basic Sciences, Calcutta, India,
for the hospitality.

\vskip 1cm


\begin{thebibliography}{99}
\frenchspacing
\bibitem{dewitt1} B. De Witt {\it Phys. Rev.} {\bf 85} (1952) 653.
\bibitem{dewitt2} B. De Witt {\it Rev. Mod. Phys.} {\bf 29} (1957) 377
\bibitem{edwards} S. F. Edwards and Y. V. Gulyaev {\it Proc. Roy. Soc.}
{\bf A279} (1964) 229.
\bibitem{rivers} R. J. Rivers {\it Path integral methods in Quantum
Field Theory}, Cambridge monographs in Mathematical Physics, 1987.
\bibitem{dirac} P. A. M. Dirac {\it Lectures on Quantum Mechanics} New York
Belfer Graduate School of Sciences, Yeshiva University.
\bibitem{falck} N. K. Falck and A. C. Hirshfeld {\it Eur. J. Phys.}
{\bf 4} (1983) 5.
\bibitem{kleibook} H. Kleinert {\it Path Integrals in Quantum mechanics
Statistical and Polymer Physics}, World Scientific 1995.
\bibitem{kleinert} H. Kleinert and S. V. Shabanov {\it Phys. Lett.}
{\bf A232} (1997) 327.
\bibitem{girotti} A. Foerster, H. O. Girotti abd P. S. Kuhn {\it Phys. Lett.}
{\bf A 195} (1994) 301.
\bibitem{marinov} M. S. Marinov {\it Phys. Rep.} {\bf 60} (1980) 1.
\bibitem{omotesato} M. Omote and H. Sato {\it Prog. Theor. Phys.}
{\bf 47} (1972) 1367.
\bibitem{asaa} A. Saa {\it Class. Quantum Grav.} {\bf 14} (1997) 385.
\bibitem{orfeu} O. Bertolami {\it Phys. Lett.} {\bf A154} (1991) 225.
\bibitem{lee} T. D. Lee {Particle Physics and Introduction to Field Theory}
Contemporary Concepts in Physics, vol. 1, Harwood Academic Publishers, 1990.
\bibitem{rb} R. Banerjee hep-th/9607199.
\bibitem{mn} T. Maskawa and H. Nakajima {\it Prog. Theor. Phys.} 
{\bf 56} (1976) 1295.
\bibitem{kapoor} A. K. Kapoor {\it Phys. Rev.} {\bf D29} (1984) 2339.
\bibitem{gh} H. Grundling and C. A. Hurst hep-th/9712052

\end{thebibliography}
\end{document}